# The physical limit of logical ">" operation

*PAN Feng[1], ZHANG Heng-liang[2], QI Jie[1]*

(1. School of Information Science and Technology, Donghua University, Shanghai 200051, China;
2. School of Power and Mechanical Engineering, Wuhan University, Wuhan 430072, China)

**Abstract**: In this paper two connected Szilard single molecule engines (with different temperature) model of Maxwell's demon are used to demonstrate and analysis the logical compare operation. The logical and physical complexity of ">" operations are both showed to be kTln2. Then this limit was used to prove the time complexity lower bound of sorting problem. It confirmed the proposed way to measure the complexity of a problem, provided another evidence of the equivalence between information theoretical and thermodynamic entropies.

## Ⅰ. INTRODUCTION

There are already many ways of measuring complexity in the field of complex systems. These measures of complexity are divided into four categories by Lloyd [1]. Computational complexity is a particularly important objective to us in complex systems [2]. Certain amount of resources is necessary in order to solve any given problems. For instance, space (memory), time and energy are all essential to run an algorithm on a computer. We proposed a definition to measure the complexity of a problem: the minimum energy required to solve it [3]. This measure was based on entropy reduction (negentropy) between the entropy before the problem was solved and after, which is interesting because the definition of information is also negentropy.

As computers are firstly physical systems: the laws of physics dictate what they can and cannot do [4,5]. Or in other words, information is physical [6]. The physics of information told us everything in the universe is doing computation. Physical objects might be thought of as being made of information [7]. The position and velocity of an atom in a gas register information. As pointed out by Fredkin and Toffoli [8,9], each atomic collision performs AND, OR, NOT, or COPY operations on suitably defined input and output bits. In this way, the second law of thermodynamics is a statement about information processing: the underlying physical dynamics of the universe preserve bits and prevent their number from decreasing [7].

Discussion on the relationship between energy and information can be traced back to the Maxwell demon which is proposed by Maxwell in 1867. The demon is a construct that can distinguish the velocities of individual gas molecules and then separate hot and cold molecules into two domains of a container, after that the two domains will have different temperature [10]. The result seems to contradict the second law of thermodynamics. Maxwell demon can be expressed by entropy which is the most influential concept to arise from statistical mechanics. The definition of entropy is: $S = k \cdot lnW$. In which $k$ is Boltzmann constant, $W$ is the number of all possible microstate (complexion) which give the same macrostate.

To exorcise the Mawell's demon, at the beginning, it's generally believed that the demon needs to cost energy in the measurement, eg. in order to locate the molecules it needs to illuminate them [11]. However, Rolf Landauer [12], the pioneer of the physics of information and Charlies Bennett [13] show that the measurement process can, in principle, be performed without energy expenditure. In fact, this measurement can be performed by a CNOT gate which is reversible with no energy cost [14]. They finally succeeded in finding the right answer which is now the famous Landauer's principle [12]: the results of the measurement must be stored in the demon's memory. The demon will need to erase his memory for new measurement as his memory is finite. There is energy dissipation associated with this erasure.

Shizume[15] and Piechocinska [16] provides examples and proofs of Landauer's principle in the domains of both classical and quantum mechanics. This erasure principle has been verified experimentally recently [17]. It can be represented in several ways.

Physical dynamics preserve information: Any process that erases a bit in one place must transfer that same amount of information somewhere else.

The connection between information and energy: In order to acquire a bit information you need to consume at least kTln2 energy.

Szilard devised in 1929 a one-molecule engine model which captured the essence and the significance of information in the thermodynamics [18]. After that it became a standard model in the physics of information research field [10]. The Landauer principle of information erasure can be explained by the standard Szilard engine model [19].

The classical computer, which is based on irreversible gates (AND gate, OR gate), is intrinsically dissipative. In contrast, reversible computer is based on reversible gates (Toffoli gate, Fredkin gate [8]). As was shown by Bennett [20], in principle, there is no energy dissipation in reversible computer.

"The required computation can be performed, print the result and run the computation backward, again using reversible gates, to recover the initial state of the computer without any energy consumption".

As we understand it, the above reversible computer is not truly, completely reversible because you still have to record the computation result in memory ("print the result" in the above description) which need energy consumption according to Landauer principle. In fact, this specific amount of energy consumption is very important and is exactly what we use to measure the complexity of a problem [3].

In previous studies, the emphasis was still put on the basic logic gates which constitute the computer [21]. But these AND, OR, NOR gates are too subtle and useless in the analysis of practical problems (eg: sorting problem). In this paper we'll concentrate on a simple problem: logical compare operations (eg: ">"). These operations are the most basic logical operation in any computer language and in the same time an important operation to many practical algorithms. We'll try to measure its logical and physical complexity and show their equivalence.

This paper is organized as the follows. In Sec. Ⅱ we focus on the logical complexity of ">" problem. In Sec. Ⅲ we map this problem into Szilard model and derive its physical complexity. In Sec. IV we apply this limit to prove the time complexity lower bound of the sorting problem. Section V is devoted to summary and discussions.

## II.LOGICAL COMPLEXITY OF ">" PROBLEM

In the following we'll take the ">" operation as the example of compare operation. The method used by Landauer [12] was chosen.

Table 1. The truth table of one bit ">" operator

| First bit | Second bit | ">" result | State |
|---|---|---|---|
| 0 | 0 | 0 | A |
| 0 | 1 | 0 | B |
| 1 | 0 | 1 | C |
| 1 | 1 | 0 | D |

Table 1 is the truth table of one bit ">" operator. As for the entropy reduction of one bit ">" operator, the problem space is 3 bit. There are eight possible initial states, in which the result is random distributed because the operation is not performed (before) and in thermal equilibrium they will occur with equal probability. The result will be determined after the computation. How much entropy reduction will occur before and after the operation is performed? The initial and final machine states are shown in Table 2. State A, B, C and D occur with a probability of 1/4 each.

Table 2. One bit ">" operator which maps eight possible states onto only four different states

| Before | | | After | | | Final |
|---|---|---|---|---|---|---|
| bit | bit | result | bit | bit | result | state |
| 0 | 0 | 0 | 0 | 0 | 0 | A |
| 0 | 0 | 1 | 0 | 0 | 0 | A |
| 0 | 1 | 0 | 0 | 1 | 0 | B |
| 0 | 1 | 1 | 0 | 1 | 0 | B |
| 1 | 0 | 0 | 1 | 0 | 1 | C |
| 1 | 0 | 1 | 1 | 0 | 1 | C |
| 1 | 1 | 0 | 1 | 1 | 0 | D |
| 1 | 1 | 1 | 1 | 1 | 0 | D |

The initial entropy was

$$S_i = k\ln W = -k\sum \rho \ln \rho = -k\sum \frac{1}{8}\ln\frac{1}{8} = 3k\ln 2$$

The final entropy was

$$S_f = -k\sum \rho \ln \rho = -k \cdot 4 \cdot (\frac{1}{4}\ln\frac{1}{4}) = 2k\ln 2$$

The difference $S_1 - S_2 = k \ln 2$

If the ">" operator was tested against two *M* bit string, it can be easily deduced that the entropy reduction will still be $k \ln 2$. To our surprise, the complexity of ">" operator has nothing to do with the length of string been tested. This result can be easily expanded to other basic compare operators: "<","<=",">=","=","!=". Their complexities are all $k \ln 2$ that have nothing to do with the length of string been tested.

In the same time, our method of reasoning gives no guarantee that this minimum is in fact, even in principle, physical achievable. It should be mentioned that the algorithm with the above minimum energy consumption is still not available now. For classical computer we may never find such algorithm (because it has to perform the operation one bit by one bit). But we believe in the possibility in other kind of computers and try to map it into physical world with Szilard single molecule engine model.

## Ⅲ. THERMAL COMPLEXITY OF ">" PROBLEM

The Szilard engine consists of a one-dimensional cylinder, whose volume is *V*, containing s single-molecule gas and a partition that works as a movable piston. The operator (a demon) of the engine inserts the partition into the cylinder, measures the position of the molecule, and connects to the partition a string with a weight at its end. These actions by the demon are optimally performed without energy consumption [13,19].

In order to get the physical resource limit of ">" operation, we must first map the logical ">" operation into physical world. The main idea is to use two connected Szilard single atom engines with different temperature. The demon's memory is also modeled as a single-molecule gas in a box with a partition in the middle. The one bit information (0 or 1) is represented by the position (left or right) of the molecule in the box.

The following is the protocol to get the information of comparison result (see Fig.1). In the beginning, the two molecules move freely over the volume *V*.
**Step 1:** There were two same cylinders (their length are both 2*L*) each with a molecule in it. But the velocity ($v_1, v_2$) of the two molecules are different. As

$\frac{1}{2}mv^2 = \frac{3}{2}kT$ and the Szilard engine is one dimension.

$$\frac{1}{2}mv_1^2 = \frac{1}{2}kT_1 \qquad \frac{1}{2}mv_2^2 = \frac{1}{2}kT_2 \qquad (1)$$

The ">" can be considered as been performed between two different velocity, kinetic energy or temperature in this way.
**Step 2:** The demon measures the location of the molecule. Then two partitions are inserted at the center of two cylinders. The first molecule was put on the right side of the first cylinder and the second molecule was put on the left side.

There is only one particle, the probabilities that particle belongs to the two sides are both 1/2, the corresponding entropy is: $S_1 = k\frac{1}{2}ln2 + k\frac{1}{2}ln2 = k\,ln2$. After separation, suppose the particle was limited to the left side. The probability that particle belongs to the left domain is 1 and the probability belongs to the right domain is 0. The corresponding entropy is: $S_2 = k1ln1 = 0$. According to the second law of thermodynamics, the minimum energy required by the demon is $T(S_1 - S_2) = kT\,ln2$. This is exactly the Landauer principle. The demon has to "borrow" $kT_1\,ln2 + kT_2\,ln2$ energy from out system in this step.

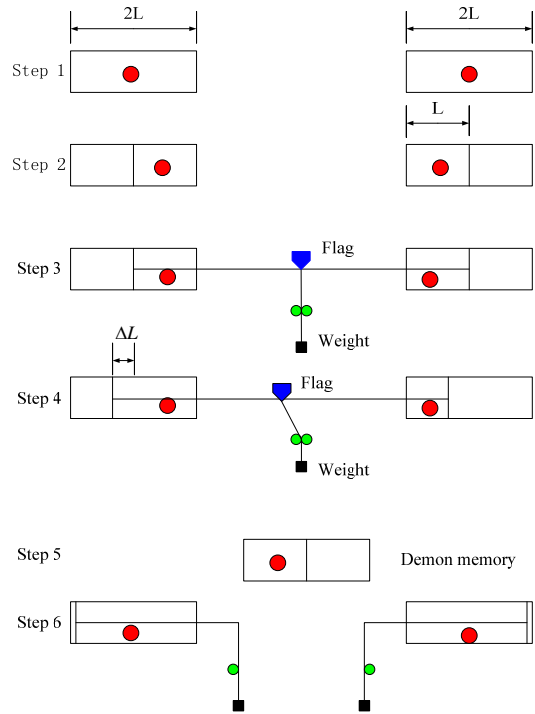

FIG.1:   The protocol to get the information of comparison result

**Step 3.** The demon connects to the two partitions a string with a weight at its middle. The weight is very small and the string won't shrink or elongate. There is a flag in a certain position on the string.

**Step 4.** Then release it, the two molecules will push the partitions. Suppose $v_1 > v_2$, the two partitions will move toward left until new balance is established. The distance moved is $\Delta L$. According to the ideal gas function (one molecule version), $pV = kT$ and in the one dimension case, The $V$ will be in direct proportion to length. Before the movement of two partitions, we'll have

$$p_1 L = kT_1, \quad p_2 L = kT_2 \tag{2}$$

After the movement of two partitions (new balance),

$$p(L+\Delta L) = kT_1, \quad p(L-\Delta L) = kT_2 \qquad (3)$$

Then

$$\Delta L = \frac{T_1 - T_2}{T_1 + T_2} L \qquad (4)$$

In the process of step 4, the weight is elevated. The amount of work extracted by the engine is:

$$\int_{L_1}^{L_2} p\, dV = \int_{L_1}^{L_2} \frac{kT}{V} dV = kT \ln \frac{L_2}{L_1} \qquad (5)$$

The work done by the left engine is $kT_1 \ln \frac{L+\Delta L}{L}$, while the work done by the right engine is $kT_2 \ln \frac{L-\Delta L}{L}$.

**Step 5:** The demon measures the location the flag and records the measurement outcome in his memory. $v_1 > v_2$ (or $T_1 > T_2$) is true. The demon will cost energy $kT \ln 2$ according to Landauer principle.

**Step 6:** Connect to the two partitions two different strings each with a weight at its end then release it. The left engine will do work $kT_1 \ln \frac{2L}{L+\Delta L}$ and the right engine $kT_2 \ln \frac{2L}{L-\Delta L}$. The sum of the work done by the two engines (step 4 plus step 6) will be:

$$kT_1 \ln \frac{L+\Delta L}{L} + kT_2 \ln \frac{L-\Delta L}{L} + kT_1 \ln \frac{2L}{L+\Delta L} + kT_2 \ln \frac{2L}{L-\Delta L} = kT_1 \ln 2 + kT_2 \ln 2 \qquad (6)$$

Which equals to the energy "borrowed" by the demon in step 2 and can be "pay back" to the out system.

In the whole process, the total energy consumed by demon to fulfill the ">" operation is $kT \ln 2$ (step 5).

## IV. THE APPLICATION OF ">" LIMIT

We will use this physical limit to prove the time complexity lower bound of sorting problem as an application. This lower bound has been proved in computer science long ago [2], but we'll rethink this problem from physical view.

Suppose there are *N* particles with different velocities in one container. The demon knows the information of their velocity but not the information of their positions. The container was first divided into $N$ equal parts. The demon then separates these $N$ particles into the $N$ left-to-right positions according to their velocities. If these $N$ particles were treated as $N$ variables, and their velocities as the corresponding variable's values, through this mapping the demon can be treated as a special purpose computer that deals with the problem of sorting.

The entropy before separation is:

$$S_1 = k \ln W_1 = k \ln N^N = kN \ln N \qquad (7)$$

The entropy after separation is 0 and the minimum energy required to solve this problem was $kTN \ln N$.

There are two kind of sorting algorithm: Comparison and not Comparison. In any specific comparison sort algorithm, the main structure is usually circulation. The following statements (in C language) or like are expected to execute in one cycle.

…
```
if (x[i]>x[j])                    1
    {
        temp=x[i];                2
        x[i]=x[j];                3
        x[j]=temp;                4
    }
```
…

According to Landauer principle, the three assignment operation (determinacy operation, sentences 2,3,4) need not consume energy in principle. The one need energy to operate is the branch statement (sentence 1 which is a kind of uncertainty operation). The operation it performs is ">", with the lower bound of energy consumption as $kT \ln 2$

Then the time complexity lower bound of sorting is:

$$\frac{kN \ln N}{k \ln 2} = N \log_2 N \qquad (8)$$

As for not comparison sorting algorithm (eg. radix sort), their time complexity is $O(N)$, but it needs extra data structure called bucket (at least $N$ buckets) compared with comparison sort, each bucket needs at least $log_2 N$ bit. In each of $N$ cycle, the radix sort need to write one bucket, this will cost $kT \ln 2 \cdot log_2 N = kT \ln N$ energy according to Landauer principle. Even radix sort cannot exceed the above energy lower bound. We show the equivalence (energy cost) between comparison and not comparison sorting algorithm, which has not yet been discovered in computer science.

## V. SUMMARY AND DISCUSSION

In this paper we verified the definition of problem complexity been proposed with a specific, simple ">" problem, provided new and profound understanding of computation complexity from the physics of information view.

We show how to measure the complexity of compare operation and demonstrated only the theoretical feasibility in physics. The idea we propose to explain the thermal limit of ">" operation, its key point is first to generate a signal (the result of ">" operation) by using two connected Szilard engine. The important conclusion we got is that this signal generation process is energy free. The second process is to measure the above signal and record the result in one bit.

There are certainly other different ways to realize this compare operation physically and some definitely can be experimental verified like the Landauer principle.